\documentclass[aip,preprint]{revtex4-1}

\usepackage{graphicx}
\usepackage{amssymb}
\usepackage{amsmath}
\usepackage{natbib}
\usepackage{color}

\begin{document}

\preprint{Internal repport}

\title{Low frequency sound attenuation in a flow duct using a thin slow sound material} 

\author{Y. Aur\'egan}\email{yves.auregan@univ-lemans.fr}
\affiliation{Laboratoire d'Acoustique de l'Universit\'e du Maine, UMR CNRS 6613
Av. O Messiaen, F-72085 LE MANS Cedex 9, France}
\author{Maaz Farooqui}
\affiliation{ASU Sound and Vibration Laboratory, Faculty of Engineering, Ain Shams University 11517 Cairo, Egypt}
\author{Jean-Philippe Groby}
\affiliation{Laboratoire d'Acoustique de l'Universit\'e du Maine, UMR CNRS 6613
Av. O Messiaen, F-72085 LE MANS Cedex 9, France}

\begin{abstract}

We present a thin subwavelength material that can be flush mounted to a duct and which gives a large wide band attenuation at remarkably low frequencies in air flow channels.  To decrease the material thickness, the sound is slowed in the material using folded side branch tubes. The impedance of the material is compared to the optimal value, which differs greatly from the characteristic impedance. In particular, the viscous and thermal effects have to be very small to have high transmission losses. Grazing flow on this material increases the losses at the interface between the flow and the material.

\end{abstract}

\maketitle 

\begin{figure}[h]
\includegraphics[width=0.9\columnwidth]{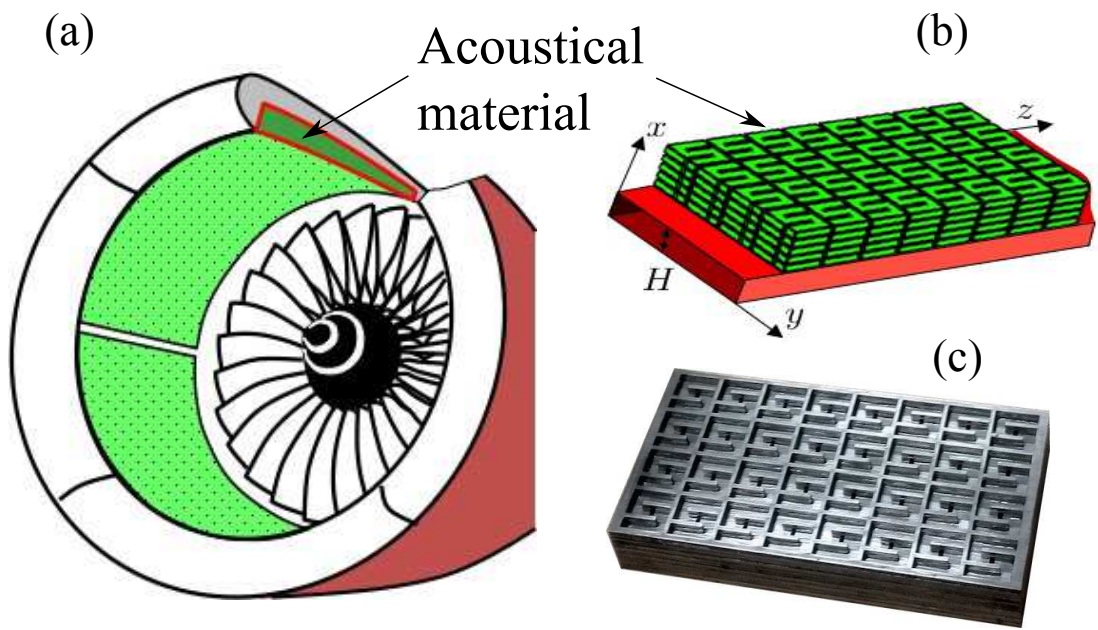}%
\caption{\label{fig_1} \textsl{(a) Sketch of a turbofan engine intake where new thin materials acting at low frequencies are needed due to the growth of fan diameter with thinner nacelles and to the reduction of the rotation speed\cite{envia2011emerging}. (b) Thin slow sound material in a duct. (c) Picture of the manufactured TSSM where the first perforated plate is removed.}}%
\end{figure}

Ducts with airflow are used in many systems, such as ventilation in vehicles and buildings, gas turbine intake/exhaust systems, aircraft engines etc.  The associated generation of unsteady flow inevitably leads to noise problems. At low frequencies, this noise is very difficult to suppress or mitigate with devices whose thickness is much smaller than the sound wavelength. There is a need for innovative acoustic materials efficient at low frequencies and able to cope with the stringent space constraints resulting from real applications (from turbofan engines, see Fig. \ref{fig_1}, to ventilation in high-rise buildings).

Recent advances in metamaterials have inspired many new designs for wave absorption by thin materials, such as absorbing membrane\cite{ma2014acoustic}, labyrinthine acoustic metamaterial\cite{liang2012extreme}, coplanar spiral resonators\cite{cai2014ultrathin}  and coherent absorbers\cite{romero2014perfect}. 
Recently the use of slow sound material has been proposed \cite{groby2015use}. By decreasing the effective compressibility in a tube\cite{auregan2015slow}, the effective sound velocity can be drastically reduced and therefore the material thickness to the same extent.  Slow sound propagation currently attracts interest in acoustic research and  has been studied both in sonic crystals\cite{cicek2012slow} 
and in one-dimensional (1D) systems with a series of detuned resonators\cite{santillan2014demonstration,theocharis2014limits}. 

The past researches on sound absorption with new materials were focusing on absorbing panels or on the reflection at tube ends. In these cases, the best attenuation is obtained when the acoustic impedance matches the characteristic impedance of air  $Z_0=\rho_0 c_0$ where $\rho_0$ is the air density and $c_0$ is the sound velocity (hereinafter, all the impedances are normalized by $Z_0$). The situation is very different in ducts with airflow. To avoid energy losses, the material had to be embedded in the wall, flush mounted and with a smooth interface to avoid any flow disturbance. The acoustic waves are no longer normal to the material. If the material is locally reacting (i.e. if the pressure and the normal velocity at the wall are linked by an impedance), the optimal impedance at frequency $f$ for an infinitely long material is the Cremer optimal impedance\cite{tester1973optimization} given by 
$Z_c=(0.91 - 0.76\mathrm{i})\; 2 f H/c_0$ 
in a two-dimensional (2D) waveguide of height $H$ . In general, this impedance differs significantly from the normal incidence one  ($Z_{ni}=1$) and the solutions developed in latter case can be ineffective when they are flush-mounted to the wall of an airflow duct.

Moreover, a mean flow is generally present in ducts and its effect on the acoustic behavior of in-duct systems has to be studied. For example,  the attenuation effects of noise barriers made with a sonic crystal can be completely destroyed by impinging air flow\cite{elnady2009quenching} and the mean flow can lead to supplementary propagative modes\cite{auregan2015slow} or to instabilities\cite{auregan2008experimental, auregan2014experimental}.

In this letter, we analyze the acoustic behavior of a thin slow sound material (TSSM) located on the sidewall of a rectangular duct with and without flow. This material has been optimized to provide a large attenuation in the low frequencies range ($\sim$ 600 Hz) despite of its small thickness ($\sim$ 27 mm), which is significantly sub-wavelength. The thickness of a conventional material (quarter wavelength resonators) would be 140 mm to be efficient in the same frequency range.

\begin{figure}[h]
\includegraphics[width=\columnwidth]{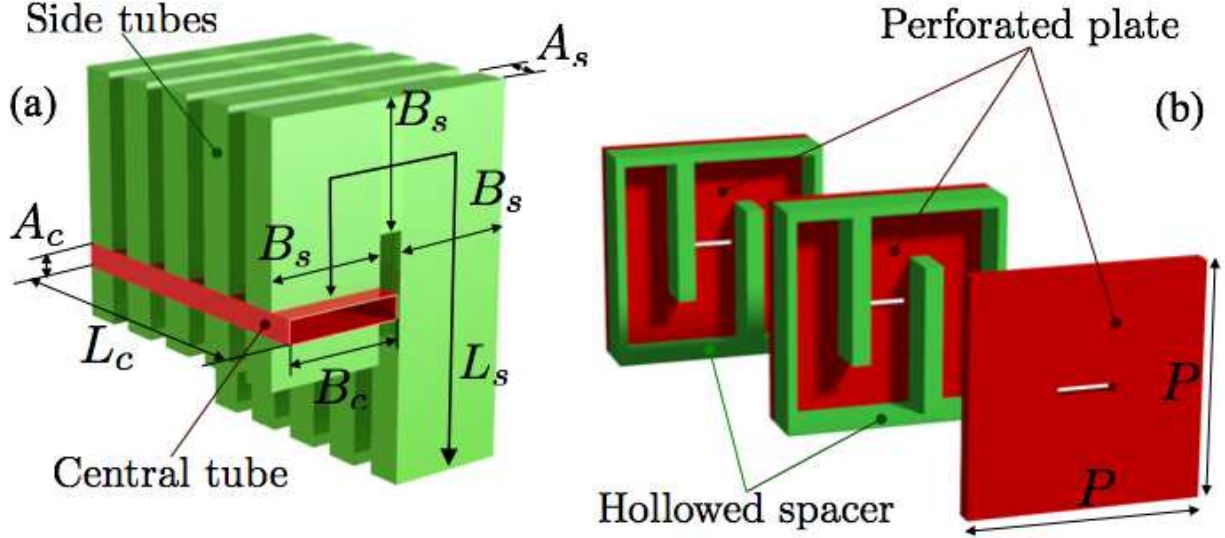}%
\caption{\label{fig_2} \textsl{(a) Description of the material where the bottom side tubes are cut for a greater clarity. (b) Implementation of the TSSM with perforated plates  and hollowed plates. }}%
\end{figure}

\begin{figure}[h]
\includegraphics[width=\columnwidth]{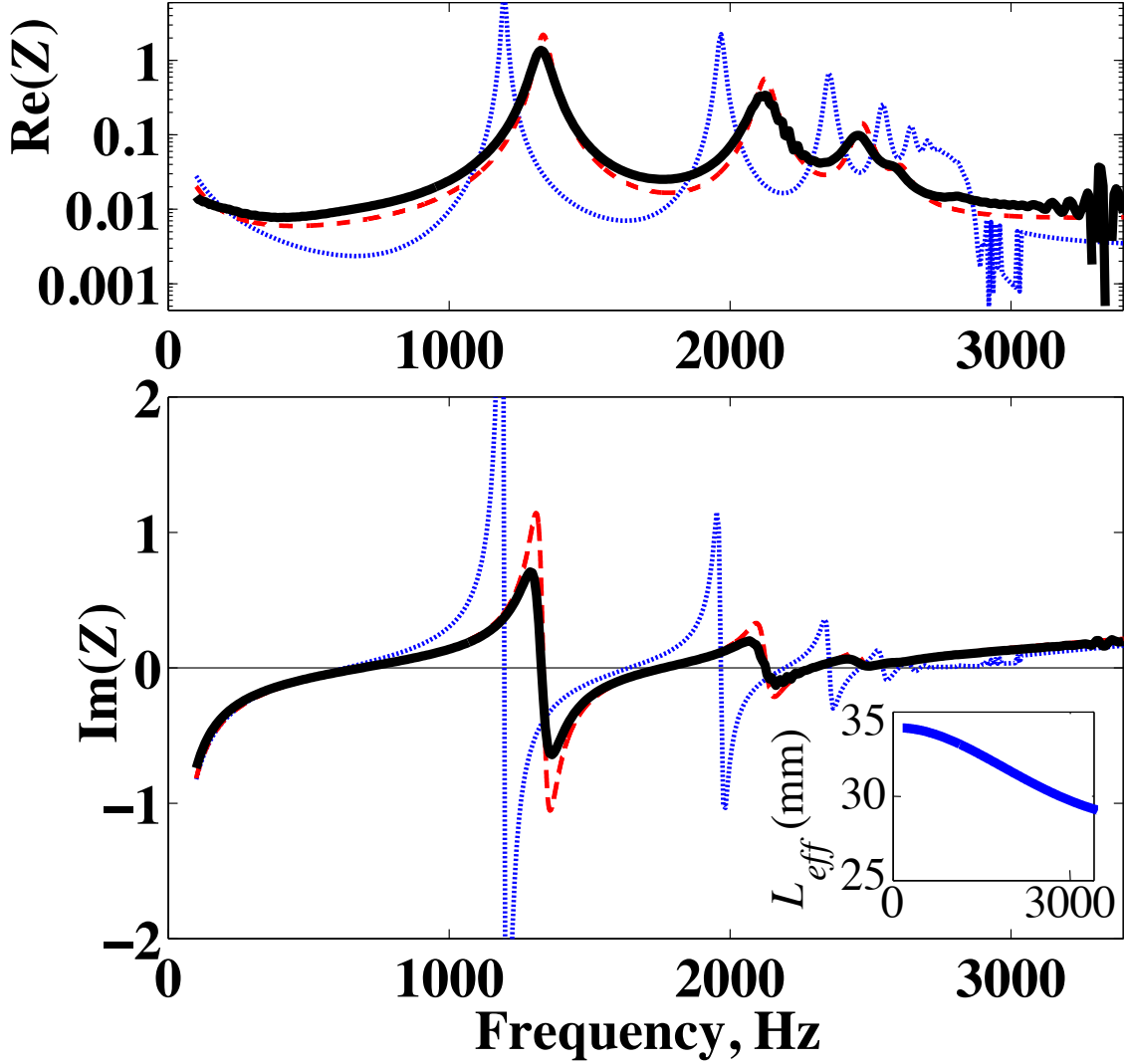}%
\caption{\label{fig_3} \textsl{Real and imaginary part of the TSSM entrance impedance. The continuous black line is a direct measurement of the impedance of one cell, the dotted blue line is the impedance computed by the continuous model and the dashed red line is the result of the lumped discrete model.  The inset gives the effective length of the side tubes computed by FEM.  }}%
\end{figure}

The studied TSSM is composed of 4 $\times$ 8 identical cells. The dimension of each cell is $P\times P \times L_c$ with $P$ = 24 mm and $L_c$ = 27 mm (see Fig. \ref{fig_2}). The cell is composed of a central rectangular tube of size $A_c\times B_c \times L_c$ with $A_c$=2.2 mm and $B_c$=6 mm. One end of the central tube is connected to the airflow duct and the other is rigidly closed. On both large sides of the central tube, 5 side branches are connected, each of them is composed of folded rectangular tubes of size $A_s\times B_s \times L_s$ with $A_s$=4 mm, $B_s$=6 mm and of neutral axis length $L_s$ = 34 mm. The $\Gamma$-shape of the side tubes increases the compactness of the material and has an effect on the acoustic response: The effective length of those tubes, that has been computed using a finite element method (FEM) in two dimensions (2D), is slightly changed on the frequency range of interest (see inset on Fig. \ref{fig_3}).

 The entrance impedance of one cell has been measured with an impedance sensor within which there are two microphones and a piezoelectric source (see appendix A of \citet{theocharis2014limits}). The real and imaginary parts of this impedance are displayed in thick black lines in Fig. \ref{fig_3}.

The propagation in the folded side tubes can be derived from a wide tube approximation of the classical Kirchhoff’s solution.  The wavenumber is given by $k_s=k_0(1+\Gamma_v+\Gamma_t)$ where $\Gamma_v$ and $\Gamma_t$, are complex numbers respectively related to viscous and thermal effects\cite{tijdeman1975propagation}. 
The normalized characteristic impedance of the side tube is $z_s=1+\Gamma_v-\Gamma_t$ and the entrance impedance of the folded side tubes is  $Z_s = - \mathrm{i} z_s \cot(k_s L_s)$. In the low frequency limit ($k_s L_s \ll 1$), the side tubes impedance is given by $Z_s = - \mathrm{i} (1-2 \Gamma_t)/(k_0 L_s)$ implying that the main dissipative effect is the thermal one at low frequencies. 

In the central tube, two different models can be applied. The first one is a continuous model where the side loaded tubes are substituted by an equivalent impedance applied on the sidewalls and the losses in the central tube are neglected. The propagation is governed by the Helmholtz equation: $\Delta p + k_0^2 \,p = 0$ where the pressure is searched under the form $p(x,y,f)=(c_1 \sinh(\alpha y)+c_2 \cosh(\alpha y))\exp(-\mathrm{i} k_c x)$ ($x$ is the direction of the central tube axis and $y$ is transverse, the convention $\mathrm{i} \omega t$ is adopted) with $\alpha^2=k_c^2-k_0^2$. Associated to the boundary condition in $y$ on the equivalent walls: $p=\mathrm{i}  Z_s/(\Phi k_0) \partial_y p$ at $y=0$ and $p= - \mathrm{i}  Z_s/(\Phi k_0) \partial_y p$ at $y=A_c$ where $\Phi$ is the equivalent wall porosity, this leads to the dispersion relation
\begin{equation}
\left(1-\left(\mathrm{i} \frac{Z_s}{\Phi} \frac{\alpha}{k_0} \right)^2 \right) \tanh(\alpha A_c) + 2 \frac{\mathrm{i} Z_s}{\Phi} \frac{\alpha}{k_0}=0.
\label{eq2}
\end{equation}

A low frequency limit  of Eq. (\ref{eq2}) can be found when $\alpha A_c \ll 1$ and the dissipative effects are neglected. The solution of Eq. (\ref{eq2}) is then approximated by:
\begin{equation}
k_c = \beta k_0= \sqrt{1+\frac{2\Phi L_s}{A_c}} \: k_0
\label{eq3}
\end{equation}
The sound speed in the central tube is reduced by a factor $\beta \sim 5$. Thus, the TSSM will be efficient at a frequency 5 times smaller than a classical material.￼￼￼￼￼￼￼￼￼￼￼￼￼￼￼￼￼￼￼￼￼￼￼￼￼￼￼￼￼￼￼￼￼￼￼￼￼￼￼￼￼￼￼￼￼￼￼￼￼￼￼￼￼￼￼￼￼￼￼￼￼￼￼￼ When the frequency increases,  Eq. (\ref{eq2}) is solved numerically to find $k_c$ and the sound speed in the central tube decrease from $c_0/5$ to 0 when a quarter wavelength resonance occurs in the side tubes. The entrance impedance of this continuous model is computed by $Z_c=- \mathrm{i} k_0 \cot(k_c L_c)/k_c$ and is plotted in dotted blue line in Fig. \ref{fig_3}.

\begin{figure}[h]
\includegraphics[width=\columnwidth]{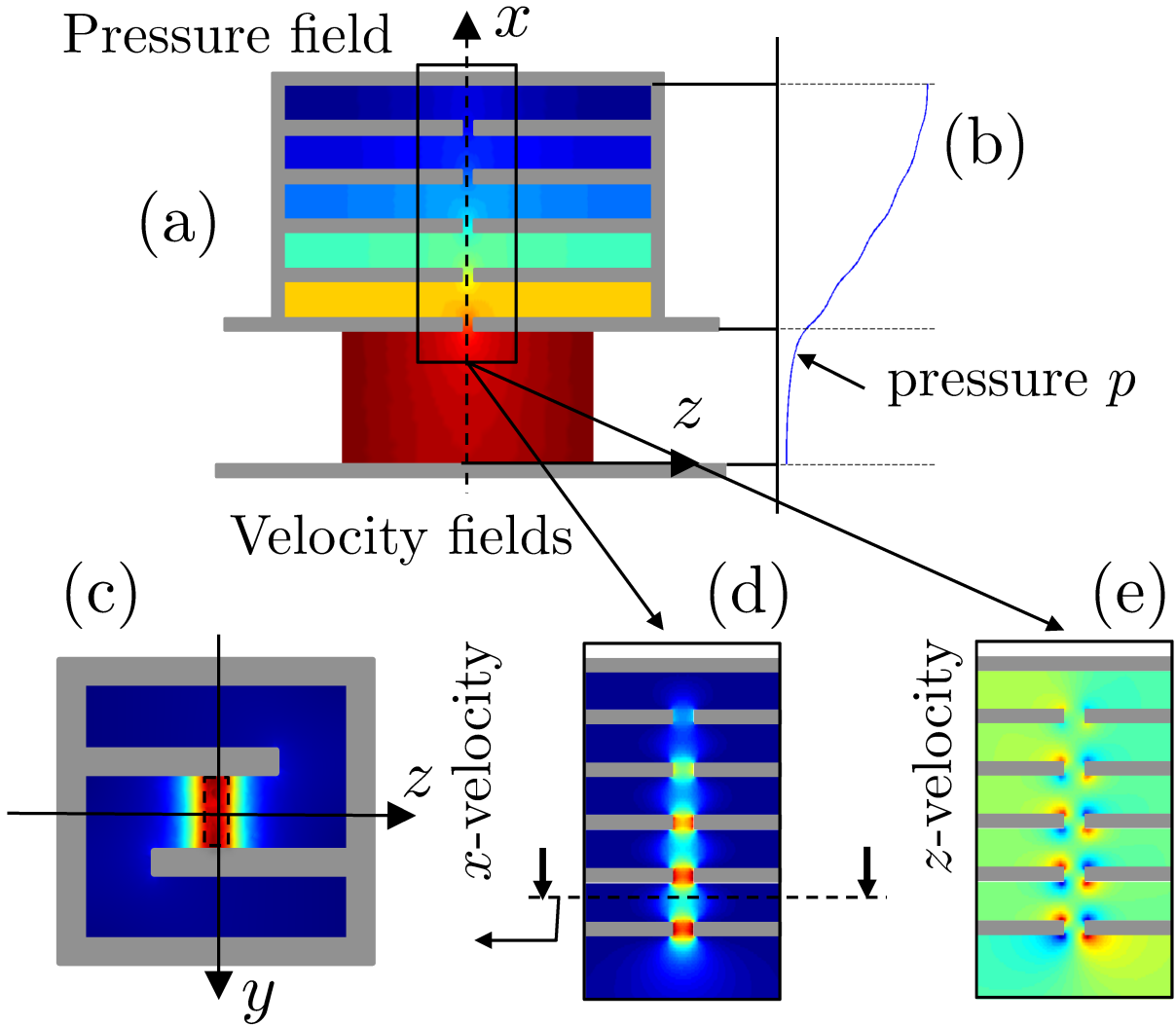}%
\caption{\label{fig_4} \textsl{FEM simulation in a perfect fluid at 800 Hz. (a) pressure field where the side branches has been unfolded, (b) shape of the pressure along the central line, (c) $x$-velocity in an $y$--$z$ plane, (d) $x$-velocity in an $x$--$z$ plane, (d) $z$-velocity in an $x$--$z$.}}%
\end{figure}

The second model is a lumped model\cite{Pierce,fang2006ultrasonic}.   It assumes that the wave propagates in the central tube with lossy hard walls apart from the central position of the loading tubes where the pressure is continuous but a part of the acoustic velocity enters in the loading tubes. In the rigid parts, the impedance is transported through the relation
\begin{equation}
Z(x_2)=\frac{Z(x_1) + \mathrm{i} \tan(k_r (x_2-x_1))}{1 + \mathrm{i}Z(x_1)\tan(k_r (x_2-x_1))}
\label{eq4}
\end{equation}
where $k_r$ is the wavenumber in the central tube accounting for the thermo-viscous losses. At the central position of the side branches, the impedance just before the discontinuity is linked to the impedance just after by
\begin{equation}
Z(x^-)= \left(Z(x^+)^{-1} +\frac{2S_s}{S_c} Z_s^{-1}\right)^{-1}
\label{eq5}
\end{equation}
Then, starting from the rigid end of the central tube where the impedance is $Z(L_c)=\infty$ and applying alternatively Eqs (\ref{eq4}) and (\ref{eq5}), the entrance impedance $Z_c=Z(x=0)$ can be found. The advantage of this method is that the complex shape of the velocity field due to the geometry complexity (see Fig \ref{fig_4}) can be accounted for by means of added masses \cite{dubos1999theory}.  In Eq. (\ref{eq5}), we replace $Z(x^+)$ with $Z(x^+)+\mathrm{i}k_0 l^{cor}_1$, $Z(x^-)$ with $Z(x^-)-\mathrm{i}k_0 l^{cor}_1$ and $Z_s$ with $Z_s+\mathrm{i}k_0 l^{cor}_2$ where $l^{cor}_1$ and $l^{cor}_2$ are the correction lengths. Taking $l^{cor}_1$=0.6 mm and $l^{cor}_2$=1.6 mm, the result of the lumped model is plotted in dashed red line in Fig. \ref{fig_3} and is in very good accordance with the measured impedance.
 
\begin{figure}
\includegraphics[width=\columnwidth]{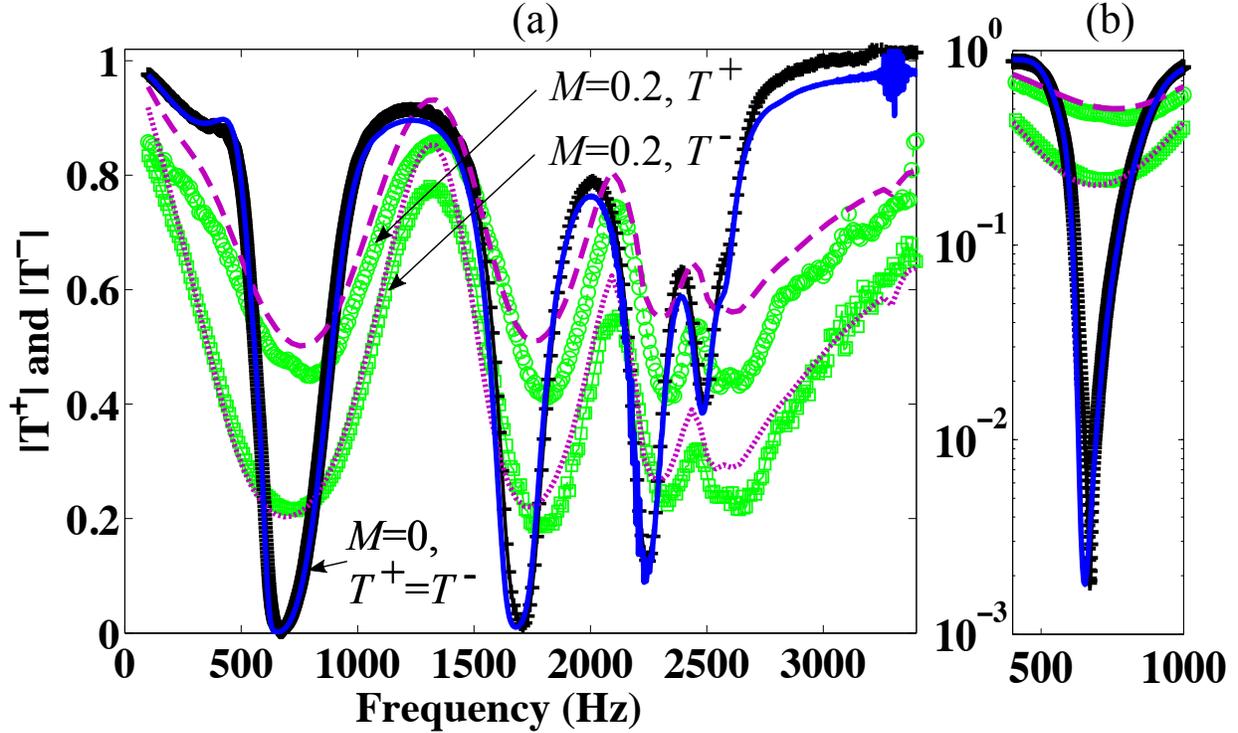}%
\caption{\label{fig_5} \textsl{Modulus of the TSSM transmission coefficients. (a) The symbols represent the measured values ($+$: without flow,  green circles: $M$=0.2, in the flow direction $T^+$, green squares: $M$=0.2, against the flow $T^-$). The lines represents the multimodal model results (blue continuous: without flow, magenta dashed: $M$=0.2, $T^+$, magenta dotted: $M$=0.2, $T^-$). (b) Zoom around the first peak in logarithmic scale.  }}%
\end{figure}

The TSSM is mounted in a rigid rectangular duct between two measurement sections, upstream and downstream. Each measurement section consists in a 100 mm $\times$ 15 mm hard walled steel duct where four microphones are flush mounted. This enables an over-determination of the incoming and outgoing waves on both sides of the TSSM.  Two acoustic sources on both sides of the system give two different acoustic states of the system and the four elements of the scattering matrix (transmission and reflection coefficient on both directions) for plane waves can be evaluated. A more detail description of the measurement technic can be found in \citet{auregan2008experimental}. The measured transmission coefficients are plotted in Fig. \ref{fig_5} in the no-flow case and in the case of a grazing flow with a mean Mach number equal to $M=0.2$.

Without flow, there is a large attenuation for three frequencies (654, 1710 and 2250 Hz). At the first peak (654 Hz), the transmitted pressure is divided by $\simeq$ 1000 (60 dB attenuation).
Around this peak, the transmission is less than 0.1 over a frequency band of 145 Hz. Accordingly, the TSSM has a very low frequency and broadband attenuation.
With flow, the attenuation is greatly reduced but occurs over a wider band.
When the wave is opposite to the flow ($|T^-|\simeq 0.2$ at 700 Hz), the attenuation is greater than when the wave and flow are in the same direction ($|T^+|\simeq 0.4$ at 700 Hz).
Mitigation vanishes in the forbidden band of TDP ($f > $ 2700 Hz) without flow, a small but non-zero attenuation exists over the entire measured  frequency range with flow. 

To better understand this behavior, the convected Helmholtz equation is solved using a rigid boundary on the lower wall ($x=0$). On the upper wall, the boundary is rigid outside of the TSSM and a Ingard-Myers boundary condition: $(k_0 - M k)^2 p =  k_0 Z_{wall} \: \partial_x  p$ is applied on the material to account for the flow effects\cite{myers1980acoustic}. 
The scattering coefficients are computed using a multimodal method\cite{Renou2011} which consists in  projecting the convected Helmholtz equation on rigid modes of the ducts in the TSSM region and in matching this solution to the propagation of the plane wave in the rigid ducts on both sides of the material. 
The key parameter is the wall impedance that can be deduced from the entrance impedance of TSSM by $Z_{wall} = (Z_c + Z_a)/\sigma$ where $\sigma$ is the percentage of open area (POA) that is computed by dividing the section of the central tube by the section of the cell ($\sigma=A_c B_c/ P^2$= 0.023). 
Without flow, the added impedance $Z_a$ account for the inertial effects in the main duct around the entrance of the TSSM, see Fig. \ref{fig_4}(d) and (e), and is equal to $Z_a= \mathrm{i}k_0 l_a$ where $l_a=1.4$ mm. 
With flow, this added length is significantly reduced and a resistance, of the order of $0.5 M$, linked to flow-acoustic interactions in the holes is added\cite{guess1975calculation} which leads to $Z_a= R_a+\mathrm{i}k_0 l_a$ with $R_a= 0.12$ and $l_a=0.5$ mm. 
The results of this multimodal model are plotted as lines in Fig.  \ref{fig_5} and  confirm that this continuous model is suitable to represent the behavior of this TSSM composed of discrete cells because the frequency is smaller than the Bragg frequency of the periodical system.

To find out if the resulting solution is optimal, the impedance $Z_{wall}$ is compared to the Cremer optimal impedance.
For the first attenuation peak (654 Hz) the resistance of the Cremer optimum (0.026) for our channel height is much smaller that the TSSM resistance (0.49). This is mainly due to the very small value of the POA that divides the entrance resistance of the TSSM (0.011) resulting from the thermal and viscous dissipation. The situation gets worse when the flow adds a resistance at the inlet holes of the TSSM and takes it away from the optimum value of resistance. In real applications, where the channel height is greater than that of our thin test duct, the optimal resistance is higher and more easily approachable. Therefore, for practical use of TSSM a balance must be reached between the thickness reduction related to the parameter $\beta$  and the POA for which an excessively low value results in a very high sensitivity to dissipation, to effects of grazing flow and to large amplitude-related effects.

The TSSM, built around the recent concept of slow sound, shows its ability to strongly reduce transmitted sounds, especially at the low frequencies, in  airflow channels.
The inherent losses of the TSSM gives rise to wide band peaks, which are widening with a grazing flow but with a reduced attenuation.
A lumped discrete approach inside the TSSM  allows an accurate modeling of the characteristics of this material.
This material proves to be a promising solution for airflow channels when space constraints and dominant low frequency noise make the classical silencers unusable.

\begin{acknowledgments} 
Y. Aur\'egan and M. Farooqui gratefully acknowledge the support from the European Union through ITN-project FlowAirS (contract  FP7-PEOPLE-2011-ITN-289352).
 \end{acknowledgments}

\end{document}